\documentclass[a4paper,11pt]{article}
\pdfoutput=1 
             
\usepackage{jheppub} 

\usepackage[T1]{fontenc} 
\usepackage{multirow}

\usepackage{subfigure}
\usepackage{float}
\usepackage{placeins}

\newcommand{\afb}{A_{\rm FB}}
\newcommand{\as}{\alpha_s}
\newcommand{\ord}{\mathcal{O}}

\newcommand{\sia}{\sigma_A}
\newcommand{\sis}{\sigma_S}

\preprint{TTK-16-51,~MITP/16-124,~ZH-TH 42/16}

\title{\boldmath The forward-backward asymmetry \\ for massive bottom quarks at the $Z$ peak \\
  at next-to-next-to-leading order  QCD}

\author[a,1]{Werner Bernreuther,\note{Corresponding author.}}
\author[a]{Long Chen,}
\author[b]{Oliver Dekkers,}
\author[c]{Thomas Gehrmann,}
\author[a]{Dennis Heisler}

\affiliation[a]{Institut f\"ur Theoretische Teilchenphysik und Kosmologie, RWTH Aachen University,\\ 52056 Aachen, Germany}
\affiliation[b]{PRISMA Cluster of Excellence and Institut f\"ur Physik, Johannes-Gutenberg-Universit\"at Mainz, \\ 55099 Mainz, Germany}
\affiliation[c]{Physik-Institut, Universit\"at Z\"urich, CH-8057 Z\"urich, Switzerland}

\emailAdd{breuther@physik.rwth-aachen.de}
\emailAdd{algeochen@physik.rwth-aachen.de}
\emailAdd{dekkers@uni-mainz.de}
\emailAdd{thomas.gehrmann@uzh.ch}
\emailAdd{heisler@physik.rwth-aachen.de}

\abstract{We compute the order $\as^2$ QCD corrections to the  $b$-quark forward-backward asymmetry  in $e^+e^-\to b{\bar b}$ collisions 
at the $Z$ boson resonance, taking the non-zero mass of the $b$ quark into account. We determine these corrections with respect to both 
the $b$-quark axis and the thrust axis definition of the asymmetry. We compute also the distributions of these axes with respect to 
 the electron beam. 
If one neglects the flavor singlet contributions to the $b$-quark asymmetry, as was done in previous computations for massless $b$ quarks,
then the second-order QCD corrections for $m_b\neq 0$ are smaller in 
magnitude than the corresponding corrections 
for $m_b=0$. Including the singlet contributions slightly increases the magnitude of the corrections. The  massive  $\as^2$ corrections
 to the $b$-quark forward-backward asymmetry slightly diminish the well-known tension between the bare $b$-quark asymmetry 
 and the standard model fit from $2.9\sigma$ to $2.6\sigma$.
 }
\keywords{QCD corrections, forward-backward asymmetry,
 bottom quarks.  }

\begin{document} 
\maketitle
\flushbottom
 
%
%
%
%
 \section{Introduction}
 \label{sec:intro}

Forward-backward asymmetries  $A_{\rm FB}^f$ are precision observables for the
 determination of the neutral current couplings of leptons and quarks
 $f$ in the reactions $e^+ e^- \to f {\bar f}$. As far as quarks are
 concerned, the most precisely known asymmetry is
 that of the $b$ quark at the $Z$ resonance, which was measured with 
  an accuracy of 1.7 percent  \cite{ALEPH:2005ab,Alcaraz:2009jr}. 
 Among the measured set of precision observables at the $Z$ pole,
 $A_{\rm FB}^b$ 
   shows a relatively large  deviation, about 2.9 $\sigma$, from
   the respective Standard Model (SM) fit. So far, it has not 
   been clarified whether this deviation is due to underestimated experimental
   and/or theoretical uncertainties or whether it is a hint of new
   physics. 
   
At a future linear or circular $e^+ e^-$ collider \cite{AguilarSaavedra:2001rg,Baer:2013cma,Gomez-Ceballos:2013zzn}, precision
determinations of electroweak parameters will again involve forward-backward asymmetries.
If such a collider will be operated at the $Z$ peak,  an accuracy of
about 0.1 percent may be reached for these observables
\cite{Hawkings:1999ac,Erler:2000jg}. 

This has motivated us to compute   $A_{\rm FB}^b$  for massive $b$ quarks produced at the $Z$ resonance
to second order in the QCD coupling $\as$.  
 The following  SM radiative corrections to the lowest-order quark forward-backward asymmetry 
 associated with quark-antiquark production in $e^+e^-$ collisions are known.
 The fully massive next-to-leading
order (NLO) electroweak and QCD corrections were determined
by \cite{Bohm:1989pb,Bardin:1999yd,Freitas:2004mn} and by
\cite{Jersak:1981sp,Arbuzov,Djouadi}, respectively.  The full
 next-to-next-to-leading
order (NNLO) QCD corrections, i.e., the contributions of  $\as^2$ to  this
asymmetry,  were recently 
 published  for the top quark in $t{\bar t}$ production above the production threshold\footnote{The forward-backward asymmetry for 
$t{\bar t}$ production at the Tevatron is also known at NNLO QCD \cite{Czakon:2014xsa}.}
in \cite{Gao:2014eea,Chen:2016zbz}.
For $b$ quarks, the order $\as^2$ corrections were 
 calculated so far only in the limit of vanishing $b$-quark mass \cite{Altarelli,Ravindran:1998jw,Catani:1999nf,Weinzierl:2006yt}.
 As pointed out in \cite{Catani:1999nf},  the  forward-backward asymmetry
 for a specific massless quark flavor $Q$ is not
 infrared  (IR)
 safe if the direction that specifies the forward and backward hemisphere
  is defined by the direction of flight of the quark $Q$ or by the
  thrust direction.  In these cases  $A_{\rm FB}^Q$ is
affected in the limit $m_Q \to 0$ by logarithmic mass
divergences $\sim \ln m_Q$. 
  These logarithmically
enhanced terms were taken into account in  \cite{Catani:1999nf} in their
   computations of $A_{\rm FB}^Q$  $(Q=b,c)$ both with respect to the
 quark and the thrust axis. In \cite{Weinzierl:2006yt} a 
 definition of  $A_{\rm FB}^Q$  based on the jet axis was given 
   that is  IR finite in the limit $m_Q \to 0$, and
 $A_{\rm FB}^Q$ was calculated for massless quarks to second order in $\as$.
  The definition used in  \cite{Weinzierl:2006yt} is an application of
 the infrared-safe definition of a flavored quark jet given in \cite{Banfi:2006hf}.

 The contributions to $A_{\rm FB}^Q$ at NNLO QCD 
   from the two-parton final state ($Q{\bar Q}$)  and from the sum of the three-
 and four-parton final states ($Q{\bar Q}g$ and $Q{\bar Q}gg$,
  $Q{\bar Q}q{\bar q}$, $Q{\bar Q}Q{\bar Q}$) to this observable are separately IR
 finite \cite{Altarelli}. The order $\as^2$ two-parton contributions to 
  $A_{\rm FB}^Q$ were computed in  \cite{Bernreuther:2006vp}. 
  Here we calculate the full order $\as^2$ QCD corrections to the $b$-quark forward-backward asymmetry, 
   both for the quark axis and the  thrust axis definition,  for 
   massive $b$ quarks to leading order in the electroweak couplings  at the $Z$ resonance.  
  
 Our paper is organized as follows. In section~\ref{sec:2} we define the forward-backward asymmetry  $A_{\rm FB}^Q$
  to order $\as^2$, unexpanded and expanded in the QCD coupling, and we briefly describe our computational 
  approach which is based on \cite{Chen:2016zbz}. In section~\ref{sec:conas2} we classify the various contributions to the 
  $b$-quark asymmetry into flavor non-singlet, flavor singlet, and triangle terms, with particular attention payed to the contribution of
   the $b{\bar b}b{\bar b}$ final state, following  \cite{Catani:1999nf}.
   Our results for the $b$-quark asymmetry to order $\as^2$ and  $m_b \neq 0$ are presented in section~\ref{secnum}. We compare our results with the QCD corrections 
    that were used in previous data analyses \cite{ALEPH:2005ab,Alcaraz:2009jr,Abbaneo:1998xt,LEPHF}. Moreover, we consider a subset of  order $\as^2$ contributions to  $A_{\rm FB}^b$
     that remain finite in the limit $m_b\to 0$ \cite{Catani:1999nf}. We compute
     the contributions of this subset for a sequence of decreasing $b$-quark masses. Extrapolating to $m_b=0$ we find agreement with the massless results of 
     \cite{Ravindran:1998jw,Catani:1999nf}. We conclude in section~\ref{sec:concl}. 
 
%
%
\section{The forward-backward asymmetry}
 \label{sec:2}
 Our computational approach applies to
  the production of any massive
quark-antiquark pair in $e^+ e^-$ collisions,
\begin{align}
e^+ e^- \rightarrow \gamma^*,\,Z^* \rightarrow Q\bar{Q} + X
 \label{eeQQX} \, ,
\end{align}  to lowest order in the electroweak couplings 
and to second order in
the QCD coupling $\alpha_s$.  To this order, the  cross section
of  the reaction  (\ref{eeQQX})   
 receives contributions from 
the two-parton $Q \bar Q$ state (at Born level, to order $\alpha_s$, 
and to order $\alpha_s^2$), the three-parton state  $Q {\bar Q} g$  (to
order $\alpha_s$ and to order $\alpha_s^2$),
and the  four-parton states $Q{\bar Q}gg$,  $Q{\bar Q} q \bar q$, and above the $4 Q$ threshold from
$Q{\bar Q}Q{\bar Q}$ (to order  $\alpha_s^2$). 
The computation of the differential cross section to order $\as^2$ was set up in \cite{Chen:2016zbz}
 within the antenna subtraction framework. We use the formulas of \cite{Chen:2016zbz} and apply it to the production
 of $b$ quarks at the $Z$ resonance.

The forward-backward asymmetry  $\afb$ for a massive quark $Q$ is defined by\footnote{For ease of notation we drop here and in the following the superscript $Q$
 in $\afb$ as we will exclusively consider $b$ quarks.}
\begin{align}
\afb = \frac{N_F - N_B}{N_F + N_B} \, ,
\label{def0afb} 
\end{align}
where $N_F$ and $N_B$ are the number of quarks $Q$ observed in the forward and backward 
 hemisphere, respectively.  Forward and backward
hemispheres are defined with respect to a certain infrared-safe axis. 
 Common choices, which we will  use in this paper,
  are the direction of flight of the heavy quark $Q$ or the direction of the oriented  thrust axis. 
 These axes are infrared and collinear safe for {\it massive} quarks; thus $\afb$ is computable
in perturbation theory. 

The asymmetry \eqref{def0afb} can be expressed in terms of the cross section for the inclusive production of 
 a massive quark $Q$~\cite{Catani:1999nf}, i.e. $d\sigma(e^+e^-\to Q+X)/dx_Q d\cos\theta$, where $\theta$ is the angle between the 
 the electron three-momentum and the axis defining the forward hemisphere and 
  $x_Q=2E_Q/\sqrt{s}$. Here $E_Q$ is the energy of $Q$ and $s$ is the squared $e^+ e^-$ center-of-mass (c.m.) energy.
   Both $\theta$ and $E_Q$  are defined in the $e^+ e^-$ c.m. frame.
  With this distribution one can define forward and backward cross sections
 \begin{align}
\sigma_F = \int_0^1 d\cos\theta \int\limits_{x_0}^1 dx_Q\frac{d\sigma}{dx_Q d\cos\theta} \, ,
 \quad 
\sigma_B =  \int_{-1}^0 d\cos\theta \int\limits_{x_0}^1 dx_Q\frac{d\sigma}{dx_Q d\cos\theta} \, ,
\label{defsiFB}
\end{align}
 and symmetric and antisymmetric cross section $\sis$ and $\sia$,
 \begin{align}
  \sis = \sigma_F + \sigma_B \, , \quad \sia = \sigma_F - \sigma_B \, .
 \label{defsiSA}
\end{align}
Here $x_0=2 m_Q/\sqrt{s}$ where $m_Q$ is the mass of $Q$.
 With \eqref{defsiSA} the forward-backward asymmetry  \eqref{def0afb} can be expressed as 
\begin{align}
\afb = \frac{\sia}{\sis} \, .
\label{defafb} 
\end{align}
Notice that above the threshold for $Q{\bar Q}Q{\bar Q}$ production the Feynman diagrams associated with this process contribute with 
 a multiplicity factor two both to $\sis$ and $\sia$ because this final state contains two quarks $Q$. 
\subsection{Unexpanded and expanded asymmetry to order $\as^2$}
\label{suse:2a}

The  forward-backward asymmetry belongs to the class of observables that can be computed
at the level of unresolved partons. 
 A number of individual terms in the following
 perturbative expansions are,  however, IR divergent and understood to
 be regulated with antenna subtraction terms as outlined in \cite{Chen:2016zbz}.

To order $\alpha_s^2$ the  symmetric and antisymmetric  cross sections
receive the following contributions from unresolved partons:
\begin{align}\label{sasbeitraege}
\sigma_{A,S} = \sigma_{A,S}^{(2,0)} +\sigma_{A,S}^{(2,1)} + \sigma_{A,S}^{(3,1)} + \sigma_{A,S}^{(2,2)} 
+ \sigma_{A,S}^{(3,2)} + \sigma_{A,S}^{(4,2)} + \ord(\as^3) \, ,
\end{align}
where the first number in  the superscripts $(i,j)$  denotes the
number of partons in the respective 
final state and the second one the order  of
$\alpha_s$. Inserting (\ref{sasbeitraege}) into (\ref{defafb}) we get
for $\afb$ to second order in $\as$:
\begin{align}\label{afbgesamt}
\afb(\as^2) = \frac{\sigma_{A}^{(2,0)} +\sigma_{A}^{(2,1)}  + \sigma_{A}^{(3,1)} + \sigma_{A}^{(2,2)}+ \sigma_{A}^{(3,2)} +
\sigma_{A}^{(4,2)}}{\sigma_{S}^{(2,0)} +\sigma_{S}^{(2,1)} + \sigma_{S}^{(3,1)}  + \sigma_{S}^{(2,2)} + \sigma_{S}^{(3,2)} +
\sigma_{S}^{(4,2)}}  =  \afb^{\rm LO}~C_2\, ,
\end{align}
where 
\begin{equation}
 \afb^{\rm LO} = \frac{\sia^{(2,0)}}{\sis^{(2,0)}} \, ,
 \label{eq:afb0}
 \end{equation}
 is the forward-backward asymmetry at Born level and $C_2$ is the  second-order QCD correction factor defined by the
  ratio on the left of this equation. The unexpanded forward-backward asymmetry at order $\as$
   is denoted by $\afb(\as)=\afb^{\rm LO}~C_1.$

A Taylor expansion of~\eqref{afbgesamt} to second order in   $\as$ gives 
\begin{align}\label{afbent}
\afb^{\rm NNLO} =&
\afb^{\rm LO}\;\left[1\;+\;A_{1} \;+\;
A_{2}\right]   \;+\; \ord(\as^3)     \, ,
\end{align}
where $A_{1}$ and $A_{2}$
are the QCD corrections of $\ord(\as)$ and  $\ord(\as^2)$, respectively.
\begin{align}
A_{1} =& \sum\limits_{i=2,3} \big[\frac{\sia^{(i,1)}}{\sia^{(2,0)}} \;-\;\frac{\sis^{(i,1)}}{\sis^{(2,0)}} \big] \, ,\label{eq: afb1} \\
A_{2} =& \sum\limits_{i=2,3,4} \big[ \frac{\sia^{(i,2)}}{\sia^{(2,0)}} \;-\;
\frac{\sis^{(i,2)}}{\sis^{(2,0)}} \big]
 - \frac{\sis^{(2,1)}+\sis^{(3,1)}}{\sis^{(2,0)}} \, A_1  \, .\label{eq:afb2}
\end{align}
The expanded NLO asymmetry will be denoted by $\afb^{\rm NLO}=\afb^{\rm LO} (1+A_1)$.

The unexpanded and expanded second-order forward-backward asymmetries \eqref{afbgesamt} 
 and \eqref{afbent} differ by terms of order $\as^3$. We will evaluate both expressions and the corresponding expressions 
  at order $\as$ in section~\ref{secnum}.

 As to the expanded version \eqref{afbent} of the forward-backward asymmetry, we recall that 
 the two-parton and the sum of the three- and four-parton contributions 
 to $A_2$ are separately infrared (IR) finite,
cf. \cite{Altarelli,Catani:1999nf,Bernreuther:2006vp}. 
 The  $Q{\bar Q}$ contribution to $A_2$ 
  is determined by the one-loop \cite{Jersak:1981sp}
   and  two-loop \cite{Bernreuther:2004ih,Bernreuther:2004th,Bernreuther:2005rw} QCD  vertex form factors 
  $\gamma^*,\,Z^* \rightarrow Q\bar{Q}$ and it was calculated in \cite{Bernreuther:2006vp} for massive $b$ and top quarks.
  The sum of the  three- and four-parton contributions to $A_2$ could be computed with any NLO method that can handle the IR divergences 
  in the three- and four-parton matrix elements individually. However, for the calculation of the numerator and denominator of the 
  unexpanded asymmetry \eqref{afbgesamt} an NNLO IR method is required. We calculate both versions of the $b$-quark forward-backward asymmetry at NNLO
  with the set-up of \cite{Chen:2016zbz}.
 
\subsection{Quark axis and thrust axis}
\label{suse:2b}
 As already mentioned above we will use both the  $b$-quark direction of flight and the oriented thrust axis
  for defining the forward and backward hemispheres.
 If the $b$-quark direction of flight is chosen  then  $\theta =\theta_b
=\angle(\vec{k}_1,\vec{p}_1)$ in eq.~\eqref{defsiFB}, where  $\vec{k}_1$ and $\vec{p}_1$ are the three-momenta of the $b$ quark and of the electron, 
 respectively, in the c.m. frame. 
 Yet, an accurate determination of the $b$-quark momentum is impeded by quark
   fragmentation and decay.
   In the past, experimental analyses often used
   the thrust axis as  reference axis. For a 
 given $n$-parton event described by a collection of final-state
 four-momenta $\{k_i\}_{i=1}^{n}$ (related by momentum conservation), 
 the thrust axis is the direction $\vec{n}_T$  that
 maximizes the thrust $T$ defined by \cite{Farhi:1977sg,Brandt:1964sa}:
\begin{equation}\label{eqThrustdef}
T=\max\limits_{\vec{n}_T}\frac{\sum\limits_{i=1}^n|\vec{k}_i\cdot\vec{n}_T|}{\sum\limits_{i=1}^n|\vec{k}_i|},\qquad |\vec{n}_T|=1.
\end{equation}
It can be shown \cite{Brandt:1978zm,Sjostrand:2006za}
 that 
\begin{equation}
\vec{n}_T\parallel\sum\limits_i\varepsilon_i\vec{k}_i,\qquad \varepsilon_i\in\{0,\pm 1\},
\end{equation}
which implies that (\ref{eqThrustdef}) is equivalent to the finite maximization problem:
\begin{equation}\label{eqThrust_eps}
T=\max\limits_{\varepsilon_i}\frac{|\sum\limits_i\varepsilon_i\vec{k}_i|}{\sum\limits_i|\vec{k}_i|}.
\end{equation}
This formula determines  $\vec{n}_T$ 
  up to a sign. Its orientation is fixed by requiring $\vec{n}_T\cdot\vec{k}_1>0$. 
  Thus, if the thrust axis is chosen as  reference axis, 
 the forward hemisphere is defined by  
$\vec{n}_T\cdot  \vec{p}_1>0$. Therefore, in this case  
 $\theta = \theta_T=\angle(\vec{n}_T, \vec{p}_1)$ in
 eq.~\eqref{defsiFB}.                
%

\subsection{Set-up of our calculation}
\label{suse:2c}

  Because we work to lowest order in the electroweak couplings, each of the various contributions $d\sigma^{(i,j)}$ 
   to the differential $b {\bar b}$ cross section to
  order $\alpha_s^2$  listed at the beginning of this section is given, at arbitrary 
   c.m. energy, by the sum of an s-channel $\gamma$ and $Z$-boson contribution  and a $\gamma Z$ interference term. 
   The $d\sigma^{(i,j)}$ have the structure
   \begin{equation}\label{eq:strucsig} 
   d\sigma^{(i,j)} = \sum\limits_{a=\gamma, Z, \gamma Z} K_a^{(j)} \: L^{\mu\nu}_a H^{(i,j)}_{a,\mu\nu} \: d\Phi_i \, .
   \end{equation}
Here  $d\Phi_i$ is the $i$-particle phase-space measure, $L^{\mu\nu}_a$ denote the lepton tensors (with the boson propagators included), and 
$H^{(i,j)}_{a,\mu\nu}$ are the antenna-subtracted, i.e., IR finite parton tensors of order $\alpha_s^j$  \cite{Chen:2016zbz}.
 Thus the Lorentz contractions and the phase-space integration in \eqref{eq:strucsig} can be done in $D=4$ dimensions.
The first index $i$ in the superscript $(i,j)$ labels the final state, i.e., $i= b{\bar b}$, $b{\bar b}g$, $b{\bar b}gg$, $b{\bar b}q {\bar q}$ $(q=u,d,s,c,b)$.
 The factors  $K_a^{(j)}$ contain the 
 electroweak couplings, the flux factor, and the $e^+e^-$ spin-averaging factor. In this work we consider unpolarized  $e^-e^+$ collisions.
 
 The electroweak neutral current couplings are
\begin{align}
v_f^Z = \,\frac{e}{2 s_W
  c_W}\,\left(T^3_f- 2 s^2_W \, e_f\right), \quad
a_f^Z = \,\frac{e}{2 s_W c_W} \,\left(-T^3_f\right),\quad
v_f^\gamma = \,e \,e_f\, , \quad a_f^\gamma = 0 \, . \label{eq:NCcoup}
\end{align}
Here $f$ denotes a quark or the electron, $e_f$ and $T^3_f$ are the charge of $f$ in units of the
positron charge $e$ and
its weak isospin, respectively, and $s_W\,(c_W)$ are the sine
(cosine) of  the weak mixing angle  $\vartheta_W$. 

  We separate each contribution $(i,j)$ on the right-hand side of  \eqref{eq:strucsig} into a  parity-even and -odd term. 
  As we work to lowest order in the electroweak couplings, these terms
  determine the cross sections \eqref{defsiSA} that are  symmetric and antisymmetric under the exchange of $b$ and $\bar b$, respectively.
   For the numerical evaluation of the $d\sigma^{(i,j)}$ we use the approach described in detail in  \cite{Chen:2016zbz}.
   In section~\ref{secnum} we consider 
  $b{\bar b}$ production exactly at the $Z$ resonance. At this c.m. energy the s-channel $\gamma$ and $\gamma Z$ interference contributions to 
   the $d\sigma^{(i,j)}$ are neglected.

\section{Contributions to order $\as^2$}
\label{sec:conas2}
 In this section we briefly discuss the various terms that contribute
 to  the $b{\bar b}$ cross section and, in particular, to the $b$-quark forward-backward asymmetry to 
 order $\as^2$. Below we shall compute the $b$-quark asymmetry also for a sequence of decreasing $b$-quark masses in order
  to compare with the massless results of  \cite{Ravindran:1998jw,Catani:1999nf}.
  For this   comparison it is useful to classify the  contributions into flavor non-singlet (NS), flavor singlet (S), and interference or triangle
 (Tr) terms. We follow here the notation and  discussion of~\cite{Catani:1999nf}. Schematically the differential cross section may be written as
 \begin{equation} \label{eq:sigdec}
  d\sigma =  d\sigma_{\rm NS} + d\sigma_{\rm S} + d\sigma_{\rm Tr} \, .
 \end{equation}
Flavor singlet and triangle contributions are present 
  only at order $\as^2$.
 The contribution of the $b{\bar b}b{\bar b}$ final state to $\afb$ deserves special attention and will be discussed in 
  section~\ref{suse:4b} below.

 \subsection{Non-singlet contributions}
 \label{suse:nonsing}
\begin{figure}[htbp]
 \begin{center}
 \includegraphics[width=12cm,height=7cm]{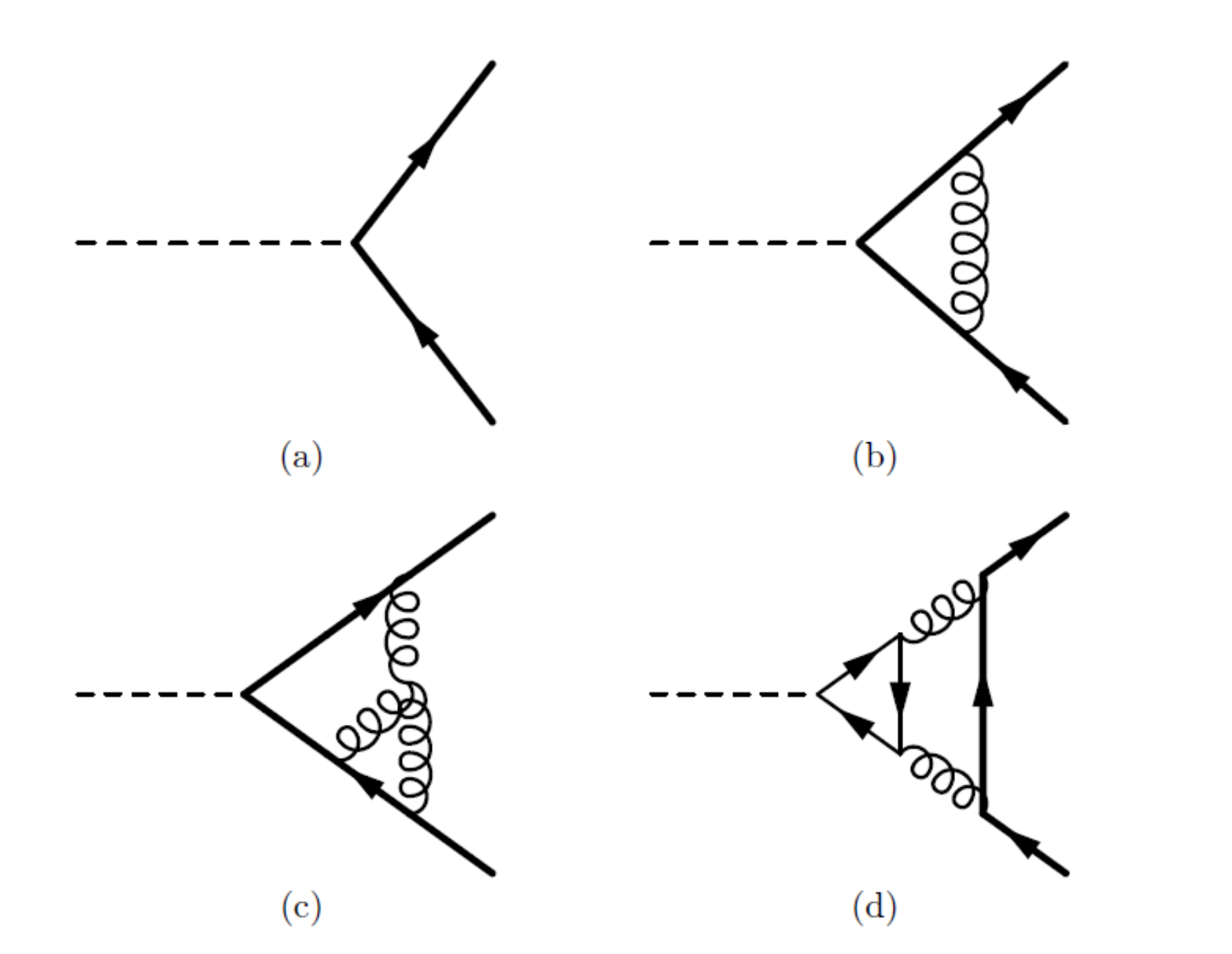}
 \caption{Examples of diagrams that contribute to the $b{\bar b}$ final state to order $\as^2$. The dashed line represents the electroweak neutral current, the 
  thick line the $b$ quark, and the thin line any of the six quarks. The triangle diagrams (d) are summed over the six quark flavors.}
\label{fig:feynman1}
 \end{center}
 \end{figure}

\begin{figure}[htbp]
 \begin{center}
 \includegraphics[width=12cm,height=4cm]{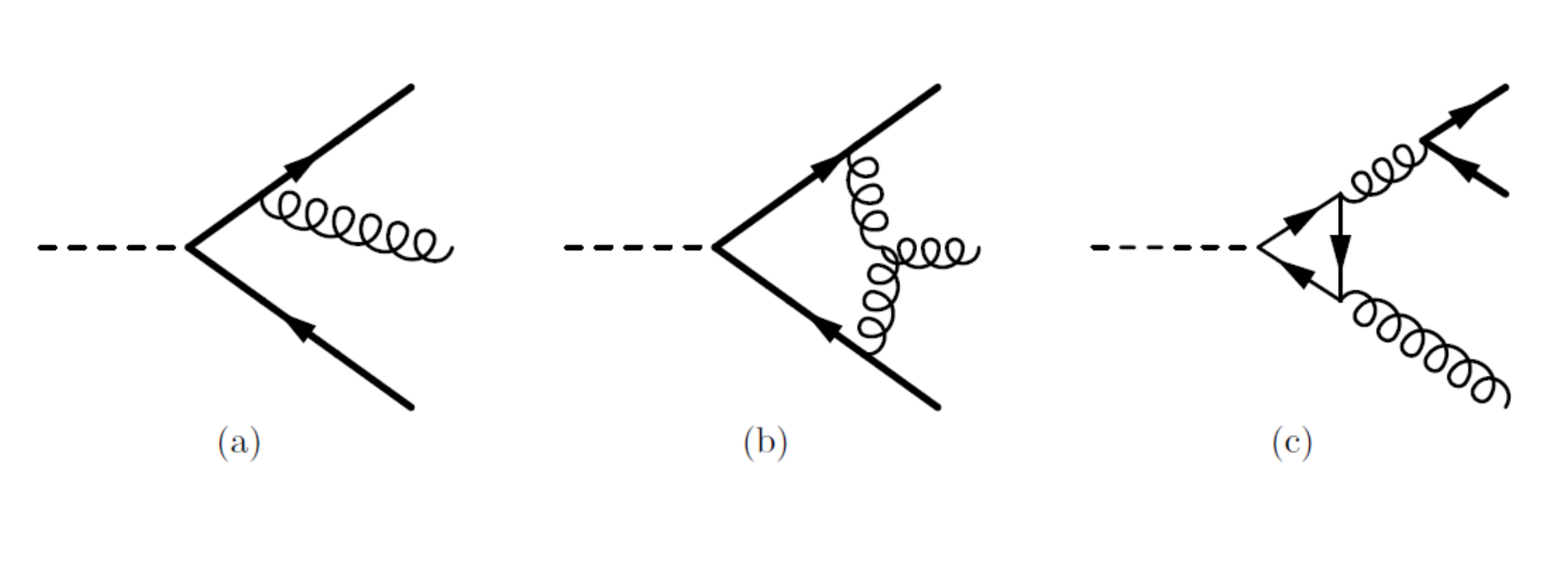}
 \caption{Examples of diagrams that contribute to the $b{\bar b}g$ final state to order $\as^2$. The assignment of the lines is as in  figure~\ref{fig:feynman1}. }
\label{fig:feynman2}
 \end{center}
 \end{figure}

\begin{figure}[htbp]
 \begin{center}
 \includegraphics[width=12cm,height=4cm]{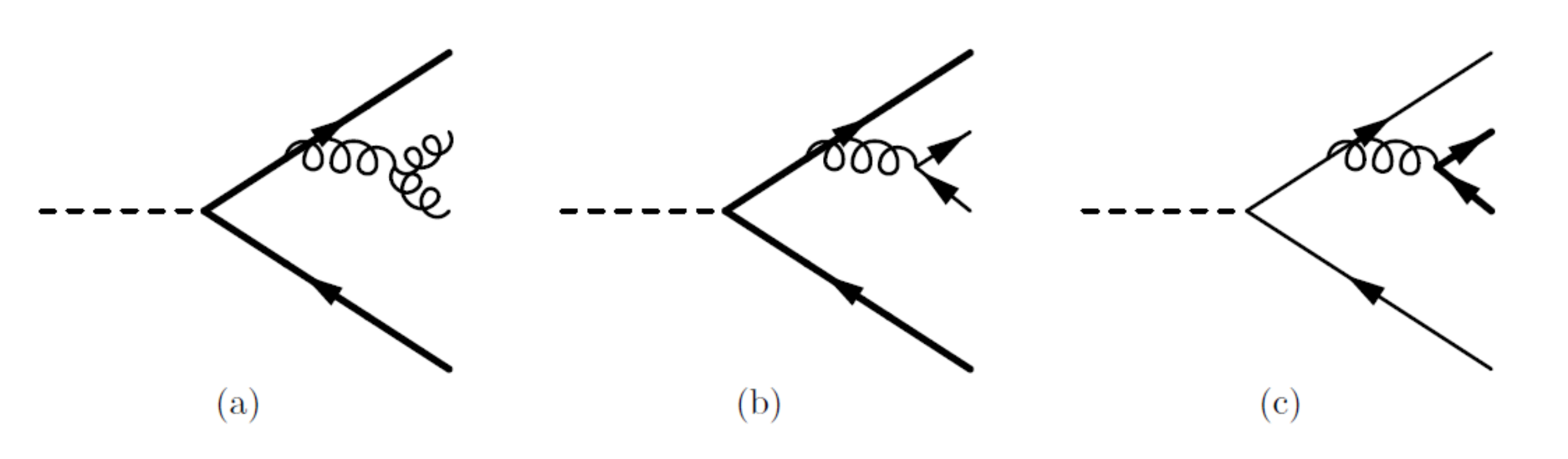}
 \caption{(a): Examples of diagrams that contribute to the $b{\bar b}gg$ final state at order $\as^2$. (b,c):  
 Diagrams that contribute to the $b{\bar b}q{\bar q}$ $(q\neq b)$ final state at order $\as^2$.}
\label{fig:feynman3}
 \end{center}
 \end{figure}

This class denotes  contributions to \eqref{eq:strucsig} and \eqref{eq:sigdec} where the electroweak current couples to the $b{\bar b}$ pair.
 As to the two-parton, i.e., $b{\bar b}$ final state: Apart from the LO and NLO diagrams (cf. figure~\ref{fig:feynman1}a and~\ref{fig:feynman1}b),  
 $d\sigma_{\rm NS}$ receives contributions of the type shown
  in figure~\ref{fig:feynman1}c. Non-singlet contributions from the three-parton final state are shown in figure~\ref{fig:feynman2}a and~\ref{fig:feynman2}b. 
   All the diagrams that correspond to the  $b{\bar b}gg$ final state (cf. figure~\ref{fig:feynman3}a) and the square of $b{\bar b}q{\bar q}$ $(q\neq b)$ diagram 
   figure~\ref{fig:feynman3}b  belong to this class, too.
    There are also contributions from the $b{\bar b}b{\bar b}$ final state, see section~\ref{suse:4b}.

 \subsection{Triangle contributions}
 \label{suse:triang}
This class involves Feyman diagrams with quark triangles, namely the interference  between the diagrams in figure~\ref{fig:feynman1}a  and~\ref{fig:feynman1}d, 
between the diagrams in figure~\ref{fig:feynman2}a  and~\ref{fig:feynman2}c,  
 and between the diagrams in figure~\ref{fig:feynman3}b  and~\ref{fig:feynman3}c.  
 The triangles in figures~\ref{fig:feynman1}d  and~\ref{fig:feynman2}c  represent a sum over all six quark flavors which couple to the respective axial current.
 We use massless $u, d, c, s$ quarks. Their contributions cancel pairwise because up- and down-type quarks have weak isospin quantum numbers $T^3_q$ of opposite sign.  The non-vanishing 
 contributions to the triangles in figures~\ref{fig:feynman1}d  and~\ref{fig:feynman2}c, that is, the differences between the $b$- and $t$-quark triangles, are ultraviolet and infrared finite. 
 The triangle contributions are non-universal corrections to the leading order  $b$-quark forward-backward asymmetry, because they involve electroweak couplings of 
  quarks $q\neq b$. 
  
 \subsection{Singlet contributions}
 \label{suse:sing}
The square of the diagrams figure~\ref{fig:feynman3}c belongs to this class. Here the $b{\bar b}$ pair is produced by the splitting of a gluon radiated off a light quark. 
Only $\sigma_S$ receives a contribution from this class, but not $\sigma_A$. There is an additional singlet contribution to $\sigma_S$ from the $b{\bar b}b{\bar b}$ final state as
 will be discussed in the next subsection.

 \subsection{Contributions from the $b{\bar b}b{\bar b}$ final state}
 \label{suse:4b}
\begin{figure}[ht]
\begin{center}
\subfigure[][Amplitudes $\mathcal{D}_{i}$, $i=1,\ldots,4$.]
{
\label{figeeQQQQ1}
\includegraphics[width=7cm]{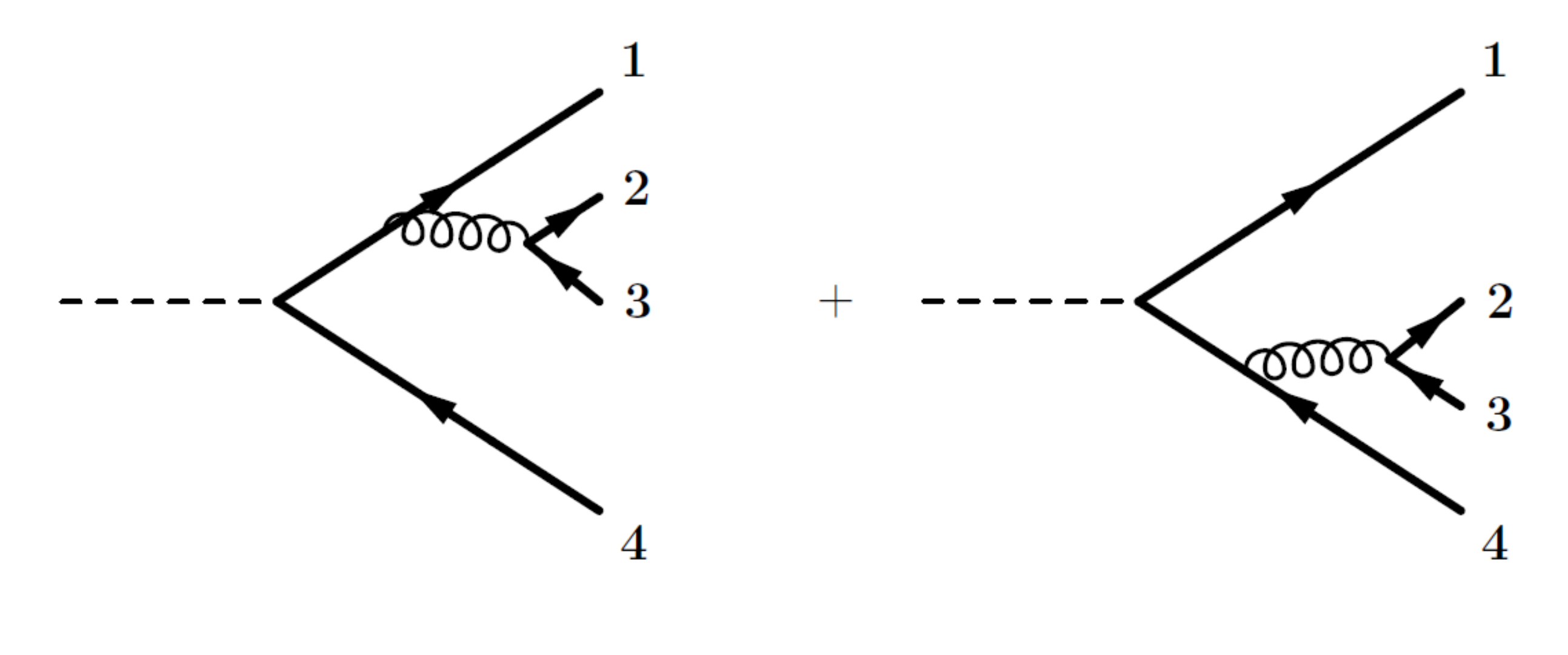}
\hspace{3ex}
}
\subfigure[][Labeling of the quark momenta]
{
\begin{tabular}[b]{c|c|c|c|c|}\label{tabeeQQQQ}
Label & $\mathcal{D}_{1}$  & $\mathcal{D}_{2}$ & $\mathcal{D}_{3}$ & $\mathcal{D}_{4}$ \\
\hline
1 & $k_1$ & $k_1$ & $k_3$ & $k_3$ \\
2 & $k_3$ & $k_3$ & $k_1$ & $k_1$ \\
3 & $k_4$ & $k_2$ & $k_4$ & $k_2$ \\
4 & $k_2$ & $k_4$ & $k_2$ & $k_4$ 
\end{tabular}
}
\end{center}
\caption{Contributions to $e^+e^- \to b
  \overline{b} \, b \overline{b}$ \label{figeeQQQQ}}
\end{figure}

 Four amplitudes $\mathcal{D}_i$, where each denotes the sum of the two diagrams  shown in figure~\ref{figeeQQQQ}, are associated with this final state because it 
contains two $b$ and two $\bar b$ quarks. 
 Therefore, as already emphasized above,  these diagrams must be counted twice in the calculation of $\sigma_S$ and $\sigma_A$.
 In our calculation of the $b$-quark forward-backward asymmetry  in section~\ref{suse:massb} we  sum these diagrams and take the square, taking into account the 
 multiplicity and statistics factor 2 and  $1/4$, respectively.
 
 Yet, for our calculation of $\afb$ for a sequence of decreasing $b$-quark masses, which is done for the purpose of comparing with the massless result of  \cite{Catani:1999nf},
 it is necessary to make a subdivision of the $b{\bar b}b{\bar b}$ term as was done in this reference. Ref.~\cite{Catani:1999nf} distinguishes between 
  (i) contributions that are identical to those of $b{\bar b}q{\bar q}$, figures~\ref{fig:feynman3}b  and~\ref{fig:feynman3}c, but with $q$ being replaced by that $b$ quark that is not triggered on, and (ii)
  genuine interference terms due to the fact that there are two indistinguishable (anti)quarks in the final state. Group (ii), which is called the E-term in \cite{Ellis:1980wv}, 
  is the color subleading part of the squared  $b{\bar b}b{\bar b}$ matrix element. In the following $D_{ij}={\rm Re}(\mathcal{D}^*_i\mathcal{D}_j)$ where, as already emphasized,  $\mathcal{D}_i$ is the sum 
  of the diagrams shown in figure~\ref{figeeQQQQ}.
  The E-term is given by the sum of the following interferences:
 \begin{align} \label{eq:Eterm1}
D_{12},\,D_{13},\,D_{24},\,D_{34} \, .
\end{align}
Ref.~\cite{Catani:1999nf} considers the $E$-term to be part of the non-singlet contributions.
 Group (i) can be partitioned into non-singlet, singlet, and triangle contributions. The singlet terms are those where the $b$ quark that is triggered on is produced by a gluon.
 By convention we assign  the momentum $k_1$ to this quark. Then the singlet contribution is given by the sum of the terms 
 \begin{align} \label{eq:4bsing}
 D_{33} \quad \text{and} \quad   D_{44} \, .
 \end{align}
 The triangle contribution is given by the sum of the terms
  \begin{align} \label{eq:4btrian} 
 D_{14} \quad \text{and}\quad D_{23} \, .
\end{align}
These are interferences between diagrams where the $b$ quark
   with momentum $k_1$  couples to the weak current and to the gluon,
   respectively. The remaining contributions to group (i) are non-singlet contributions. 

We come back to this classification in section~\ref{suse:mass0} when comparing with \cite{Catani:1999nf}.

\section{Numerical results for the $b$-quark asymmetry at the $Z$ peak}
\label{secnum}
 In this section we present our results for the $b$-quark asymmetry at NNLO QCD and  lowest order 
  in the electroweak couplings at the $Z$ resonance,
 with respect to both the $b$-quark and the thrust direction.
 As mentioned before, we put the couplings of the virtual photon to the fermions to zero.
 We use the computational framework of \cite{Chen:2016zbz}, that is, antenna-subtracted renormalized matrix elements
 with the $b$-quark mass defined in the on-shell scheme and the QCD coupling $\as$ defined in the $\overline{\rm MS}$ scheme.
 For the  $\overline{\rm MS}$ mass of the $b$ quark we take the average determined in \cite{Olive:2016xmw}: \\
\begin{equation}\label{eqmbmb}
\overline{m}_b(\overline{m}_b)=4.18 \pm 0.03~{\rm GeV} \, ,
\end{equation}
which yields the on-shell mass: 
\begin{equation}\label{eqmb}
m_b= 4.89\pm 0.04~{\rm GeV}.
\end{equation}
For the mass of the top quark that appears in the triangle diagrams figure~\ref{fig:feynman1}d  and~\ref{fig:feynman2}c  
 we use $m_t=173.34$ GeV. The $u,d,c,s$ quarks are taken to be massless.
For the  $\overline{\rm MS}$
 coupling of five-flavor QCD we take
  $\alpha_s(m_Z)=0.118$.
 The sine of the weak mixing angle,  $s_W$,  is fixed by $s^2_W=1-m_W^2/m_Z^2$. With $m_W=80.385$ GeV and $m_Z=91.1876$ GeV
 one gets $s^2_W=0.2229$.
 For computing the electroweak couplings of the quarks and the electron we use the $G_\mu$ scheme
  where the electromagnetic coupling is given by $\alpha=\sqrt{2}G_\mu m_W^2 s_W^2/\pi= 7.5624 \times 10^{-3}$ 
  with $G_\mu=1.166379\times 10^{-5}$ ${\rm GeV}^{-2}$.
  
  As mentioned above the  $b \bar b$ contribution to the second-order correction $A_2$, eq.~\eqref{eq:afb2},
   is IR finite. It was computed in \cite{Bernreuther:2006vp} with unsubtracted $b{\bar b}$ matrix elements. Here we compute the  
  $b \bar b$ contribution to $A_2$ with antenna-subtracted matrix elements. As a check of our set-up we calculated this contribution with the 
   input parameters of \cite{Bernreuther:2006vp} and found agreement with the numbers given in table~1 of this reference. We checked  also that 
   the sum of the three- and four-parton contributions to $A_2$ is IR finite.


\subsection{Massive $b$ quark, quark axis and thrust axis}
\label{suse:massb}
   
   With the values of $s^2_W$ and the bottom mass given above, the tree-level value of the $b$-quark forward-backward asymmetry at $\sqrt{s}=m_Z$ is 
   $\afb^{\rm LO}=0.1512$. The value of $\afb^{\rm LO}$ is very sensitive to the input value of $s^2_W$ but insensitive 
    to the uncertainty on $m_b$ given in \eqref{eqmb}. We are concerned here with the first- and second-order QCD corrections to the LO asymmetry.
   They are given in table~\ref{tab:afbb-exp} for the expanded version of $\afb$, both for the quark axis and the thrust axis definition,  
    for the three renormalization scales $\mu=m_Z/2, m_Z, 2 m_Z$. In this subsection we take into account all contributions 
    discussed in section~\ref{sec:conas2}.
   
 \begin{table}[tbh!]
\begin{center}
\caption{\label{tab:afbb-exp} The first- and second-order QCD correction factors defined in \eqref{afbent} - \eqref{eq:afb2} 
 to the LO $b$-quark forward-backward asymmetry 
at the $Z$ peak for the input values given in the text and for $\mu=m_Z$. The numbers in superscript (subscript) refer to the changes if
$\mu=2 m_Z$   ($\mu=m_Z/2$) is chosen.}
\vspace{1mm}
 \begin{tabular}{| c|c|c|c|c|}\hline 
 &     $1+A_1$& $1+A_1+ A_2$  & $A_1$ & $A_2$ \\ \hline 
 quark axis: & $0.9710^{+0.0028}_{-0.0034}$ & $0.9587^{+0.0026}_{-0.0028}$ & $ -0.0290$ &  $-0.0123$ \\ [2mm] \hline                             
  thrust axis: &  $0.9713^{+0.0027}_{-0.0026}$ & $0.9608^{+0.0022}_{-0.0025}$  & $-0.0287$ & $-0.0105$ \\[2mm] \hline    
 \end{tabular}
 \end{center}
 \end{table}
 
  Table~\ref{tab:afbb-exp} shows that the order $\as^2$ corrections are significant. For $\mu=m_Z$ the ratio $A_2/A_1$ is $43\%$ and $37\%$ for 
   the quark and thrust axis definition, respectively. Variation of the scale as in table~\ref{tab:afbb-exp} changes both the first-  and second-order QCD correction factors
    by about $\pm 0.003$ with respect to their values at $\mu=m_Z$. The fact that inclusion of the second-order correction term $A_2$ does not (significantly) 
     reduce the scale uncertainty is not unusual for 
     an observable that is defined as a ratio. 
    
   The first- and second-order corrections $A_1$ and $A_2$ are dominated by the contributions from the three-parton and three- and four-parton final states, respectively.
   In the limit $m_b\to 0$ the $b \bar b$ contribution to $A_1$ and the non-singlet $b \bar b$ contribution to $A_2$ vanish because the chiral
    non-singlet currents become conserved. Because $m_b/m_Z \ll 1$  these contributions to $A_1$ and $A_2$ turn out to be about two orders of magnitude smaller than the 
    three-parton, respectively three- and four-parton contributions. The $b \bar b$ triangle contribution to $A_2$ (cf. figure~\ref{fig:feynman1}d) is about one order of magnitude larger 
     than the non-singlet $b \bar b$ contribution, but an order of magnitude smaller than those from the three- and four-parton final states.
    
    We have included  in the computation of $A_2$ given in table~\ref{tab:afbb-exp} also the non-universal corrections $A_2^{\tiny{non-u.}}$ of order $\as^2$ that
   contain the vector and axial vector couplings of quarks $q\neq b$. They are significant. For instance, for the quark axis definition and $\mu=m_Z$  
   we get $A_2^{\tiny{non-u.}}=-0.00310$ which is $25\%$  of the total correction $A_2$.  This number comes about as follows. 
    The two- and the three-parton contribution to $A_2^{\tiny{non-u.}}$, that is, 
   the interference of figure~\ref{fig:feynman1}a  and~\ref{fig:feynman1}d and of figure~\ref{fig:feynman2}a and~\ref{fig:feynman2}c, is small; it is $+0.00085$ and $+0.00028$, respectively.
   The dominant part is due to the non-universal contributions from the $b{\bar b} q{\bar q}$ $(q\neq b)$ final state. While the term $\sia^{(4,2)}/\sia^{(2,0)}$ (cf. eq. \eqref{eq:afb2})
    of this correction is negligibly small, the term $\sis^{(4,2)}/\sis^{(2,0)}$ is significant. As a result the contribution of this  term to $A_2^{\tiny{non-u.}}$ is $-0.00423$.
     Notice that the value of $A_2^{\tiny{non-u.}}$ depends on the value of $s^2_W.$
   
 Next we  represent the  expanded version of the  $b$-quark asymmetry, both for the quark and the
   thrust axis,  in the form:
\begin{equation}\label{A2Qbas}
\afb^{\rm NNLO}=\afb^{\rm LO} \left[ 1-a_1 \frac{\as}{2 \pi}
       - a_2 \left(\frac{\as}{2 \pi}\right)^2 \right] \, .
\end{equation}
 Table~\ref{tab:afbb-coeff} contains  the values of the coefficients $a_1$ and $a_2$ extracted from the values of 
  $A_1$ and $A_2$ of  table~\ref{tab:afbb-exp}.

 \begin{table}[tbh!]
\begin{center}
\caption{\label{tab:afbb-coeff} The values of the first- and second-order coefficients $a_1$ and $a_2$ defined in \eqref{A2Qbas} for $\mu=m_Z/2, m_Z$,
 and $2 m_Z$.}
\vspace{1mm}
 \begin{tabular}{|c c|c|c|}\hline 
             &               & $a_1$    & $a_2$   \\ \hline 
 quark axis  & $\mu=m_Z/2$:  &  $1.544$ & $26.67$   \\ 
             &  $\mu=m_Z$:   & $1.544$ & $34.84$    \\
             &  $\mu=2 m_Z$: & $1.544$ & $43.06$     \\  \hline                             
 thrust axis & $\mu=m_Z/2$: & $1.528 $ & $21.75$   \\ 
             &  $\mu=m_Z$:  & $1.528 $ & $29.81$  \\
             & $\mu=2 m_Z$: &  $1.528 $ & $37.98$  \\ \hline 
 \end{tabular} 
 \end{center}
 \end{table}

 Monte-Carlo simulations or  measurements of the $b$-quark forward-backward asymmetry at the $Z$ peak 
  can also be compared with perturbative computations where the ratio $\sia/\sis$ is not 
  expanded. In this case the order $\as$ and order $\as^2$ correction factors $C_1$ and $C_2$ apply that are defined in \eqref{afbgesamt} and below \eqref{eq:afb0}.
  Their values are given in table~\ref{tab:afbb-unex}.

\begin{table}[tbh!]
\begin{center}

\caption{\label{tab:afbb-unex} The first- and second-order QCD correction factors 
 $C_1$ and $C_2$ defined in eq.~\eqref{afbgesamt} and below eq.~\eqref{eq:afb0}.} 
\vspace{1mm}
 \begin{tabular}{|c|c|c|}\hline 
              &     $C_1$                      & $C_2$   \\ \hline 
 quark axis:  &   $0.9722^{+0.0025}_{-0.0031}$ & $0.9594^{+0.0026}_{-0.0030}$  \\  [2mm] \hline                                
 thrust axis: &  $0.9725^{+0.0025}_{-0.0031}$ & $0.9614^{+0.0023}_{-0.0026}$  \\ [2mm] \hline     
 \end{tabular}
 \end{center}
 \end{table} 
 
 The spread between the second-order expanded and unexpanded correction factors may be viewed, in addition or alternatively to 
  scale  variations, as an indication  of the order of magnitude of the uncalculated higher-order corrections.
  The comparison of the values of $1+A_1+A_2$ and $C_2$ for fixed $\mu$ given in table~\ref{tab:afbb-exp} and~\ref{tab:afbb-unex}
    shows that both for the quark and the thrust axis definition
   the spread between these correction factors is significantly smaller than the change of these terms due to scale variations. 
   This indicates that the perturbative calculation of the $b$-quark forward-backward asymmetry is reliable.

 \begin{figure}[tbh!]
 \centering
 \includegraphics[width=7.45cm,height=11cm]{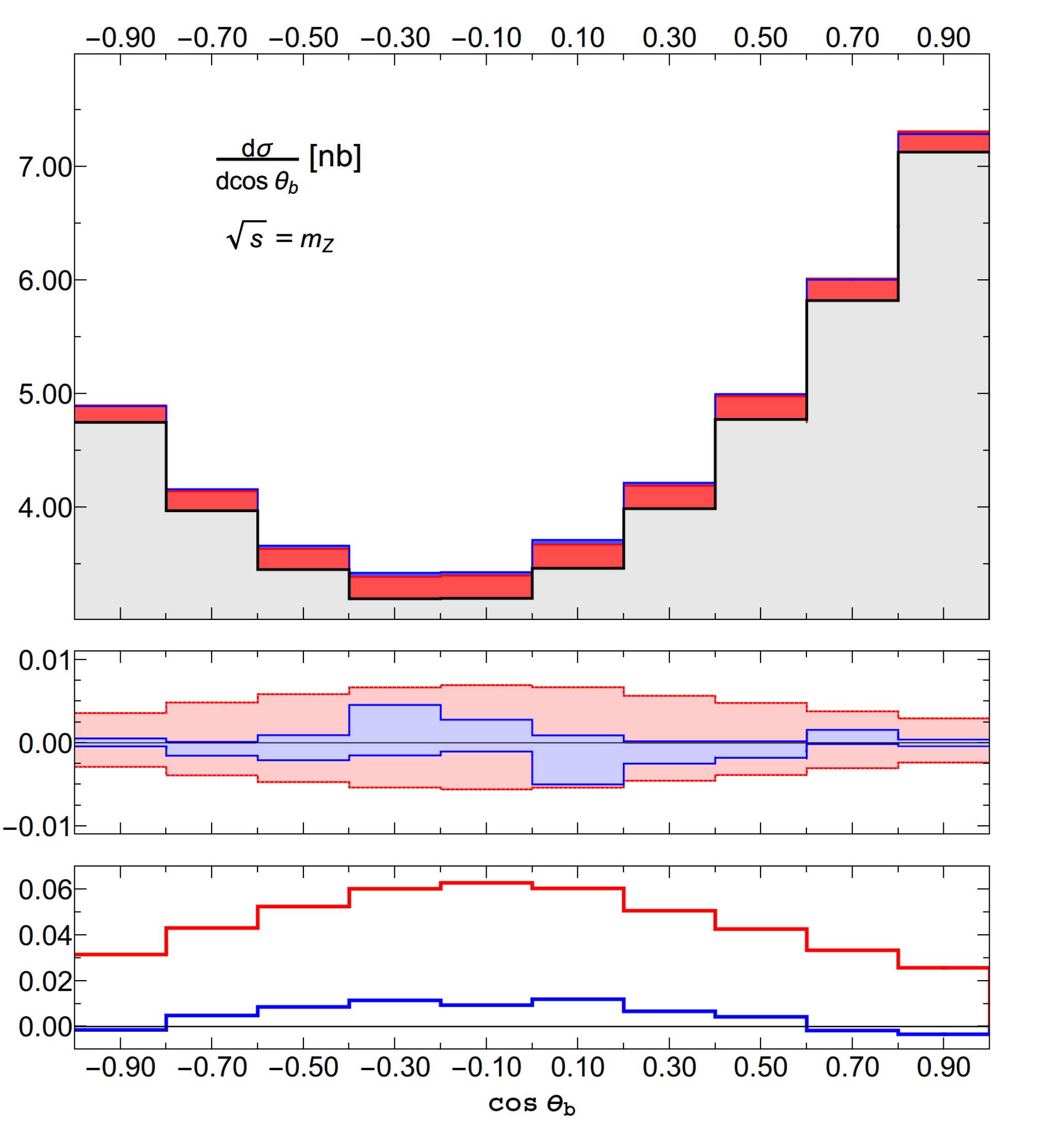}
 \includegraphics[width=7.45cm,height=11cm]{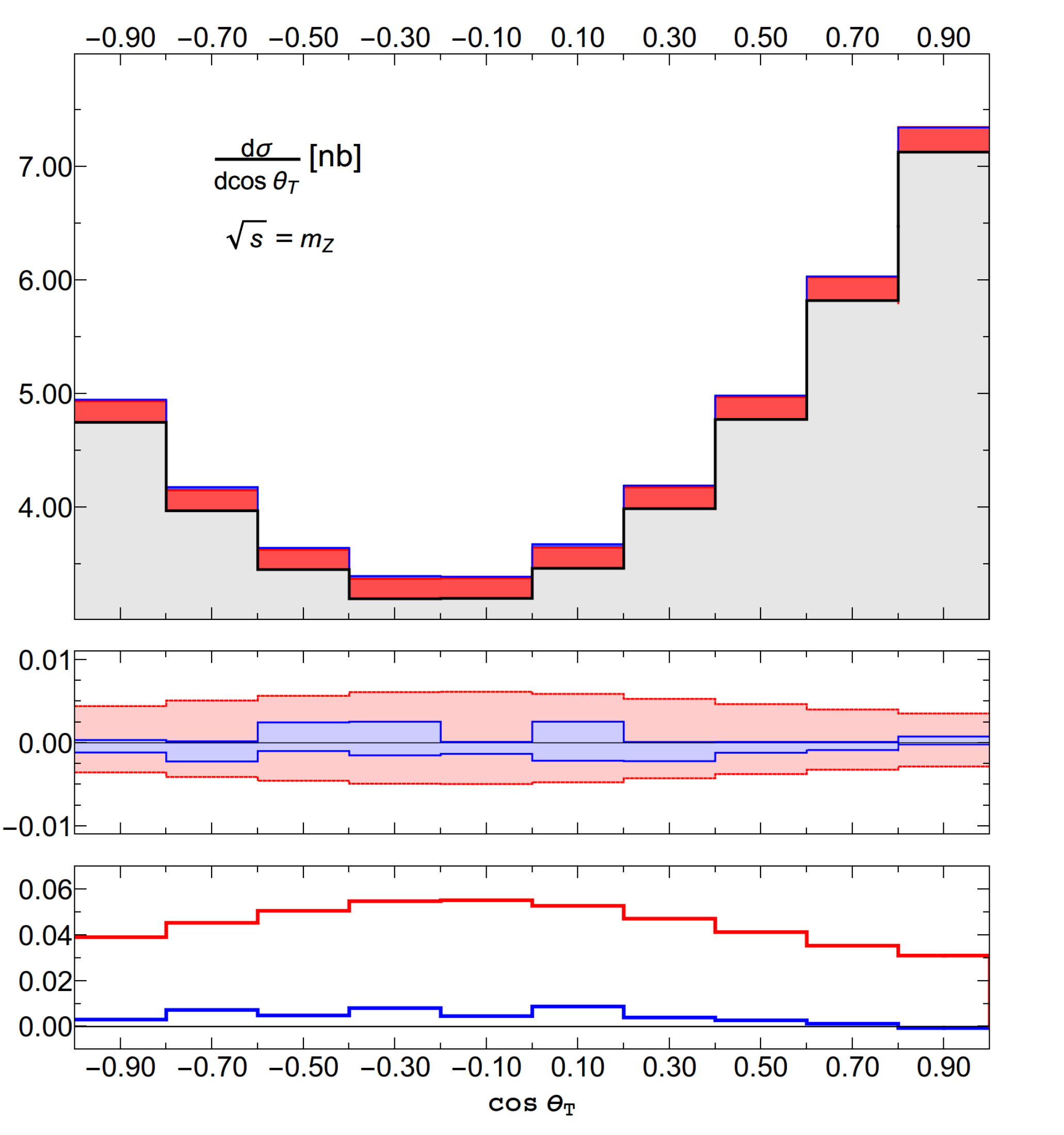}
 \caption{The distribution of $\cos\theta_b$ (plots on the left) and of  $\cos\theta_T$ (plots on the right) at $\sqrt{s}=m_Z$. The upper panels show the respective distribution
  at LO (grey), NLO (red), and NNLO QCD (blue) for $\mu=m_Z$. The panels in the middle  display the scale variations ${\rm NLO}(\mu')/{\rm NLO}(\mu=m_Z)-1$ (red band) and 
  ${\rm NNLO}(\mu')/{\rm NNLO}(\mu=m_Z)-1$ (blue band) of the first and second order QCD corrections, where $m_Z/2\leq\mu'\leq2 m_Z$. The lower panels show
   the ratios ${d\sigma_1}/d\sigma_{\rm LO}$  (red) and ${d\sigma_2}/d\sigma_{\rm LO}$  (blue) for $\mu=m_Z$. 
   }
 \label{fig:costhet}
 \end{figure}

 Finally we display in figure~\ref{fig:costhet} the distributions of  $\cos\theta_b$ and $\cos\theta_T$ to order $\as^2$, where 
 $\theta_b$ $(\theta_T)$ is the angle between the $b$-quark direction (oriented thrust direction) and  the electron beam. 
 Here we use the schematic notation $d\sigma_{\rm NLO} = d\sigma_{\rm LO}+ d\sigma_1$ and  $d\sigma_{\rm NNLO} = d\sigma_{\rm NLO} + d\sigma_2$.
 The plots show that the order $\as^2$ corrections to these un-normalized distributions are small and  these corrections reduce the scale uncertainties.

\subsection{Approaching the limit of massless $b$ quarks}
\label{suse:mass0}

Next we compute the second-order correction  to the $b$-quark forward-backward asymmetry for a sequence of decreasing values
 of $m_b$. This allows us to compare 
 with the results of \cite{Ravindran:1998jw,Catani:1999nf} obtained for $m_b=0$.
In order to conform to the  calculation of \cite{Catani:1999nf} 
 we neglect now, as was done in  \cite{Catani:1999nf},  the singlet and the triangle contributions.
 Thus we take into account only the non-singlet contributions to \eqref{eq:afb2} which we denote by  $A^{NS}_2$. 
 We recall here the classification of the various second-order contributions done in section~\ref{sec:conas2}
  that is in accord with \cite{Catani:1999nf}.
 
   It was shown in  \cite{Catani:1999nf} 
  that the second-order correction $A^{NS}_2$ becomes  singular for $m_b\to 0$ due to a logarithmic
   singularity that arises in the phase-space integration of the 
   symmetric $E$-term in the triple-collinear regions.
    As mentioned above, 
   the $E$-term is the color subleading contribution to the squared
   matrix element  of the $b\bar{b}b\bar{b}$ final state. It consists of the  interference terms listed in
    \eqref{eq:Eterm1}. It was also shown in ref. \cite{Catani:1999nf}  that $A^{NS}_2$ can be decomposed as follows:
  \begin{equation}\label{eqAfinCS0}
A^{NS}_2  = {\widehat A}_2  -  \int E_S \, ,
\end{equation}  
  where  $\int E_S$ denotes the phase-space integral over the symmetric E-term that contains a term $\propto \as^2\ln(s/m_b)$ 
   and  ${\widehat A}_2$ is finite in the limit $m_b\to 0$.
  We recall that the  $b\overline{b}b\overline{b}$ diagram contributions to
$A^{NS}_2$ are   multiplied by a factor of $2$ (cf. section~\ref{sec:conas2}).

 \begin{figure}[tbh!]
 \centering
 \includegraphics[width=7.5cm,height=6.5cm]{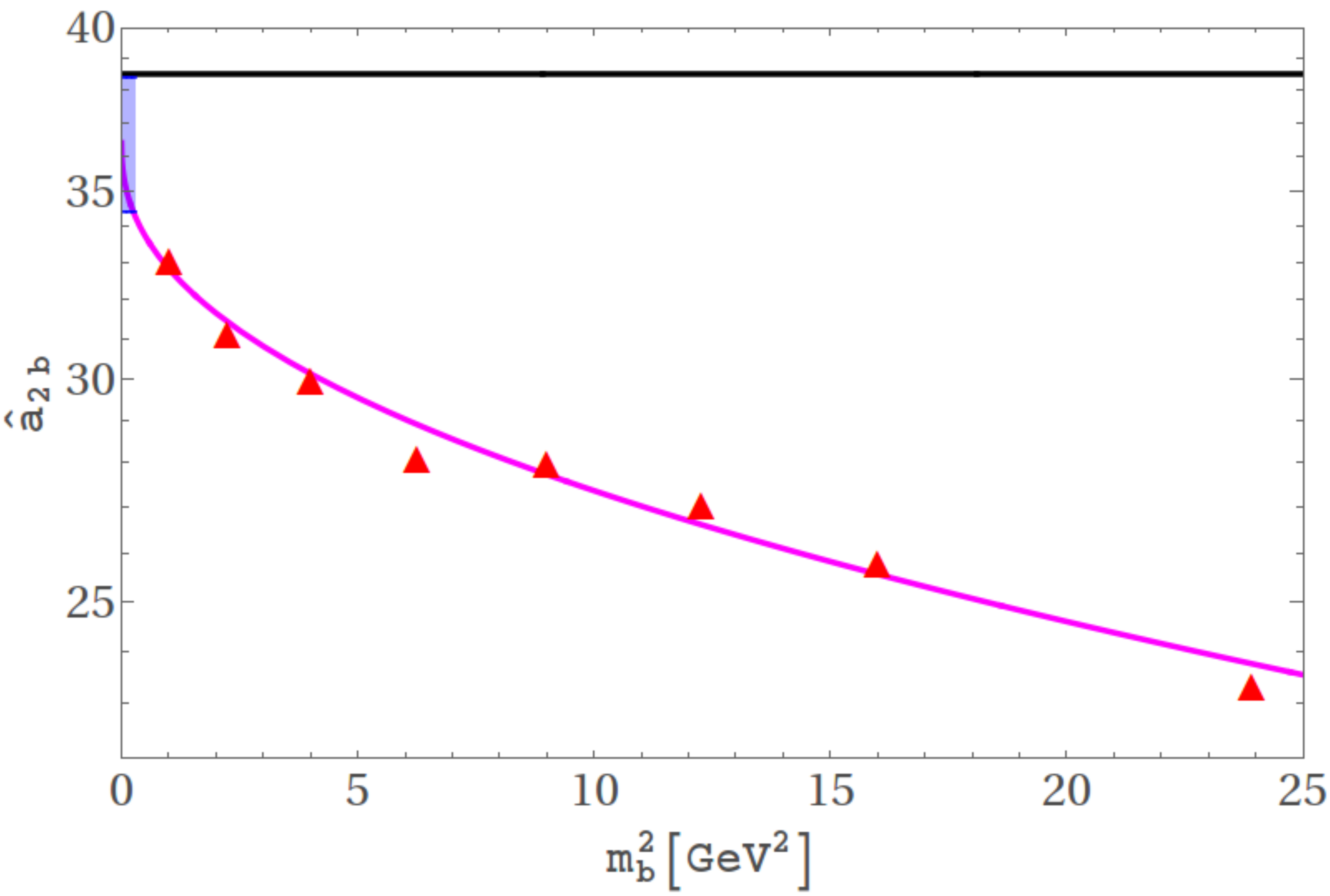}
 \includegraphics[width=7.5cm,height=6.5cm]{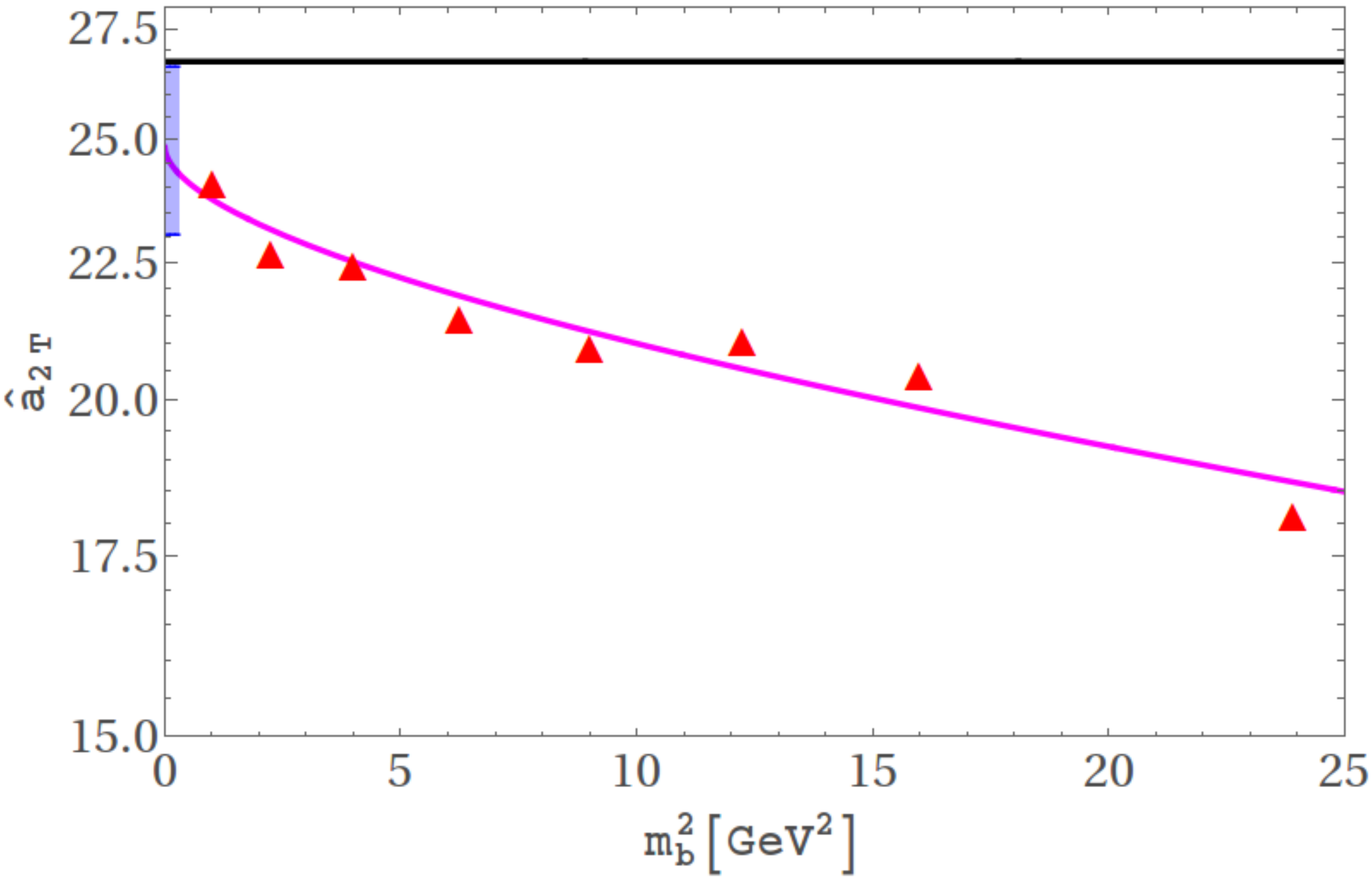}
\caption{Left plot: The solid red triangles are the values of the order $\alpha_s^2$ correction coefficients ${\widehat a}_2$ defined in \eqref{eq:a2NS} 
   for the quark axis definition 
    for a sequence of  $b$-quark mass values and $\mu=m_Z$. The solid red curve is obtained by a fit using the ansatz \eqref{eq:fitansatz}.
 The horizontal solid black line is the value for $m_b=0$ computed in
 \cite{Ravindran:1998jw,Catani:1999nf}. The shaded blue vertical line is the $1\sigma$ uncertainty of the value of ${\widehat a}_2$ at
  $m_b=0$ resulting from the fit. Right plot: same as left plot, but for the thrust axis definition of $\afb$. Here the solid black line 
 is the value of ${\widehat a}_2$ at $m_b=0$ computed in \cite{Catani:1999nf}.
   }
\label{fig:a2fit}
 \end{figure} 

    The term  ${\widehat A}_2$ was  calculated  for $m_b=0$ and for the quark axis definition 
     in \cite{Ravindran:1998jw, Catani:1999nf} and for the thrust axis definition in \cite{Catani:1999nf}.
    Here we compute  ${\widehat A}_2$, respectively the coefficient
    \begin{equation}\label{eq:a2NS}
    {\widehat a}_2 = -       \left(\frac{2\pi}{\as}\right)^2   {\widehat A}_2
    \end{equation}
    for a sequence of decreasing $b$-quark
    mass values between $m_b=4.89$ GeV and $m_b=1$ GeV. 
    We choose the renormalization scale to be $\mu=m_Z$ which was apparently also chosen
     in \cite{Catani:1999nf}. We compute ${\widehat a}_2$ both for the quark and thrust axis definition of $\afb$.
     The results are shown by the red solid triangle points in the left and right plots of figure~\ref{fig:a2fit}.
     In order to extrapolate ${\widehat a}_2$  to  $m_b=0$ we perform a fit using the ansatz
      \begin{equation}\label{eq:fitansatz}
       c_0 + c_1 z +  c_2 z\ln z^2 \, ,
       \end{equation}
 where $z=(m_b/m_Z)$. This ansatz is motivated by the leading mass terms of the NNLO corrections in the 
  limit $z\to 0$. The coefficient $c_0$ is the value of  ${\widehat a}_2$ at $m_b=0$. 
 We obtain 
 \begin{equation}\label{eq:fitres}
 \text{quark axis:} \; c_0 = 36.40 \pm 1.70\, , \qquad  \text{thrust axis:} \; c_0= 24.83  \pm 1.78\, ,
 \end{equation}
 which agree  within errors with the values
 \begin{equation}\label{eq:rvncas}
  {\widehat a}_{2b}(m_b=0) = 38.5 \, , \qquad  {\widehat a}_{2T}(m_b=0) = 26.74 
 \end{equation}
 for the quark \cite{Ravindran:1998jw,Catani:1999nf} and
  thrust axis definition \cite{Catani:1999nf}, respectively.

 If one wants to compare the size of the QCD corrections to $\afb$ for a massive and a massless $b$-quark,
  one should compare  the massless order $\as^2$ correction coefficients \eqref{eq:rvncas}
  of  \cite{Ravindran:1998jw,Catani:1999nf} with the respective coefficients shown in figure~\ref{fig:a2fit} for non-zero $m_b$, rather than
   comparing with the coefficients $a_2$ given in table~\ref{tab:afbb-coeff}, because the latter contain also singlet and triangle contributions.
   Figure~\ref{fig:a2fit} shows that both for the quark and for the
thrust axis definition, the second-order corrections are smaller in magnitude for massive
quarks than for massless ones. For $m_b=4.89~{\rm GeV}$ we obtain  
\begin{equation}\label{eq:rvm49}
  {\widehat a}_{2b}(m_b=4.89~{\rm GeV}) = 23.31 , \qquad  {\widehat a}_{2T}(m_b=4.89~{\rm GeV}) =  18.43 \, . 
 \end{equation}
The magnitude of the second-order corrections  decreases with increasing quark mass.
 This holds true also for the first-order corrections, as exemplified by comparing the 
  values of $a_1$ for $m_b=4.89$ GeV listed in table~\ref{tab:afbb-coeff} with the values 
 $a_{1b}=2$  and  $a_{1T}=1.787$ for $m_b=0$.
  This is in accord with the basic physical fact that a massive (anti)quark is
more inert than a massless one in radiating  off partons, and hence less
affected by changes of its direction with respect to the 
leading-order quark antiquark configuration.

\subsection{Discussion}
\label{suse:disc}

 The QCD corrections to the $b$-quark $\afb$  determined above (or those determined in the  
  calculations  \cite{Ravindran:1998jw,Catani:1999nf} for $m_b=0$) cannot be applied 
 directly to the analysis of experiments. 
 In the  measurements of the $b$-quark 
asymmetry reviewed in \cite{ALEPH:2005ab,Alcaraz:2009jr}, the thrust axis was used to define the forward and backward
hemispheres.  In our computation the thrust axis is defined for partonic final states,
 but the hadronization of partons causes a smearing of this axis. In addition,  a bias in the topology of the events is introduced by the experimental 
  selection and analysis method towards two-jet final states which causes additional uncertainties \cite{Abbaneo:1998xt,LEPHF}. 
  
  A proper discussion of these issues is beyond the scope of this paper. Here we only compare the QCD corrections computed above with 
   those that were taken into account  in \cite{ALEPH:2005ab,Alcaraz:2009jr,LEPHF}.  
   These analyses aimed at determining a pseudo-observable: the bare $b$-quark 
$Z$-pole asymmetry $A_{\rm FB}^{0,b}$ from  the
  measured  asymmetry      $A_{{\rm FB},{\rm exp}}^{b,T}$ with a procedure described in \cite{Abbaneo:1998xt,LEPHF}.  
    First,  $A_{{\rm FB},{\rm exp}}^{b,T}$ was corrected for QCD effects as follows (cf. \eqref{A2Qbas}):
  \begin{equation}\label{A2Qbaex}
 A_{{\rm FB},{\rm exp}}^{b,T}  =\left[ 1-a_{1T} \frac{\as}{2 \pi}
       - a_{2T} \left(\frac{\as}{2 \pi}\right)^2 \right]  (\afb^{0,b})_{\rm exp} 
       \equiv (1-C^T_{\rm QCD}) (\afb^{0,b})_{\rm exp} \, .
\end{equation}
   The  QCD corrected ``experimental''  asymmetry $(\afb^{0,b})_{\rm exp}$ was then further corrected
    for higher order electroweak corrections like photon exchange,
    $Z\gamma$ interference, and other photonic corrections (cf. for instance,
  \cite{Bardin:1999yd,Freitas:2004mn}) before a value of the bare asymmetry $A_{\rm FB}^{0,b}$ was deduced. In this way  ref.  \cite{Alcaraz:2009jr} obtained the experimental
   value for the pseudo-observable
   \begin{equation}\label{eqAFBpolalt}
A_{FB}^{0,b}=0.0992\pm 0.0016 \, ,
\end{equation}
 where the error refers to experimental and theoretical  uncertainties. The pull between \eqref{eqAFBpolalt} and the value $A_{FB}^{0,b}=0.1038$
 obtained by a combined fit to all SM precision observables  \cite{Alcaraz:2009jr} is $2.9\sigma$.
 
 The QCD correction factor defined in \eqref{A2Qbaex} and used in  \cite{Alcaraz:2009jr} was obtained as follows \cite{Abbaneo:1998xt,LEPHF}. 
 For the order $\as$ correction the value
  $a_{1T}=1.54$   was taken\footnote{This value is almost the same as the value $a_{1T}=1.53$ given in table~\ref{tab:afbb-coeff}.
 We recall that we  use a slightly larger $b$-quark mass. Moreover, we determine the thrust axis by classifying the final states according to the moduli of their 
 three-momenta rather than their energies as in \cite{Djouadi}.} that was computed in \cite{Djouadi} for $m_b=4.5$ GeV.
   For the second-order QCD  correction coefficient  the value $a^T_2(m_b=0)=23.72$ was used.
   This number is obtained  by adding to the massless result of  \cite{Catani:1999nf} (cf. \eqref{eq:rvncas}) the two-loop $b\bar b$ triangle contribution. (The sign 
    of this contribution to $a^T_2$ is opposite to that of \eqref{eq:rvncas}.) The QCD correction factor determined in \cite{LEPHF} is
   $(1-C^T_{\rm QCD})=0.9646 \pm 0.0063$ where the  error includes estimates of hadronization effects.
   Our thrust axis correction factor $(1+A_1+ A_2) = 0.9608 \pm 0.0025$  given in table~\ref{tab:afbb-exp}, 
    where the error is due to scale uncertainties only, agrees with that factor within the uncertainties. Our central value is smaller 
     than $0.9646$ by  $0.4\%$. Our correction changes the value of the pseudo-observable $A_{FB}^{0,b}$ to $0.0996\pm 0.0016$.
       Thus the pull  between  $A_{FB}^{0,b}$ and the SM fit cited above is slightly decreased, namely from $2.9\sigma$ to $2.6 \sigma$.

    We recall that our value of the QCD correction factor  was obtained by taking into account all 
    second-order QCD contributions discussed in section~\ref{sec:conas2}. If one neglects the singlet contributions, that is, if one uses the value of 
    ${\widehat a}_{2T}(m_b=4.89~{\rm GeV})$  given in \eqref{eq:rvm49} and adds the $b{\bar b}$ and $b{\bar b}g$ triangle contributions, we
     get for the thrust axis correction factor $(1+A_1+ A_2) =  0.9659 \pm 0.0023$. This value is not significantly larger than the correction factor used in \cite{LEPHF} and cited above.

%
%
%
\section{Conclusions}
\label{sec:concl}

We have computed the second-order QCD corrections to the $b$-quark forward-backward asymmetry 
in $e^+e^-\to b{\bar b}$ collisions at the $Z$ boson resonance. The mass of the $b$ quark was fully taken into account. 
We have determined these corrections  both for using the  quark and the thrust axis in defining the forward and backward hemisphere. 
We have computed also the distributions of these axes with respect to 
 the electron beam.
 The complete order $\as^2$ corrections to the $b$-quark asymmetry, that is, the sum of the flavor non-singlet, flavor singlet, and triangle contributions
 are significant; they amount to  $43\%$ and $37\%$ of the order $\as$ corrections for 
   the quark and thrust axis definition, respectively. If one neglects the singlet contributions, as was done in 
    previous calculations for massless $b$ quarks \cite{Ravindran:1998jw, Catani:1999nf}, then 
    the second-order QCD corrections for $m_b\neq 0$ are smaller in magnitude than the corresponding corrections
    for $m_b=0$. This is expected on general physical grounds. 
  We have also demonstrated that by decreasing the value of the $b$-quark mass we can approach with our computational set-up 
  the massless order $\as^2$ results of \cite{Ravindran:1998jw, Catani:1999nf}.

As emphasized above our results cannot be applied directly to the analysis of existing measurements of the $b$-quark asymmetry. We have compared the magnitude of the
second-order QCD corrections for $m_b\neq 0$ with those  used in previous analyses that deduced a bare $b$-quark asymmetry $A_{\rm FB}^{0,b}$ from the measured one. 
If one takes into account the complete massive order $\as^2$ correction then the value of the bare asymmetry \eqref{eqAFBpolalt} increases slightly, which reduces the pull
 between this pseudo-observable and the value of the standard model fit from $2.9\sigma$ to $2.6\sigma$.
 
 As a future application of our computational set-up one may consider the determination of the order $\as^2$ corrections to the forward-backward asymmetry for two $b$-jet final states, 
  for which one expects a decrease of the magnitude of the QCD corrections.

%
%
%
%
\section*{Acknowledgment}
 L. Chen acknowledges  support by a scholarship  from the China Scholarship Council (CSC).
  D. Heisler was supported  by Deutsche Forschungsgemeinschaft through Graduiertenkolleg GRK 1675. 
%
%

%
%
\end{document}